\documentclass[aps,twocolumn,prl,showpacs,floatfix]{revtex4}
\usepackage{graphicx}
\newcommand{\be}{\begin{equation}}\newcommand{\bea}{\begin{eqnarray}}
\newcommand{\ee}{\end{equation}}\newcommand{\eea}{\end{eqnarray}}
\begin{document}

\title{
Self-avoiding knots}

\author{A.Yu. Grosberg} \affiliation{
Department of Physics, University of Minnesota, Minneapolis, MN 55455 }

\begin{abstract} Scaling arguments are used to analyze the size
of topologically constrained closed ring polymer with excluded
volume.  It is found that there exists a finite range of polymer
thickness (excluded volume) in which self-avoidance is unimportant
and polymer swelling compared to the Gaussian size is entirely due
to the topological constraints.
\end{abstract}

\pacs{61.25.Hq, 82.45.-h, 82.20.Wt %\hfill cond-mat ??????? (2002)
}

\maketitle

The ability to knot is the most obvious and the least understood
property of long strings, such as ropes, threads, spaghetti, flux
lines, or polymers.  Knots in DNA have been directly observed
\cite{Cozarelli,Wang}. The compelling, albeit circumstantial,
proof of their importance can be seen in the fact that Nature had
undertaken to design special cell machinery (called topological
enzymes) spending energy to simplify DNA topology \cite{Rybenkov}.
% There are reports that cells die if their DNA is too tightly
% knotted \cite{???}.
Knots and other topological constraints are also of great
importance in equilibrium and dynamics properties of polymer
materials, such as melts, gels, networks, etc.  Despite this ample
motivation, current understanding of polymer topology is very
limited.

Apart from mathematical theory of knot invariants \cite{Kaufman},
the study of knots in polymers was dominated for a long time by
computer simulations \cite{Maxim,Muthukumar,Deguchi,Deutch}. The
incorporation of sophisticated topological invariants to compute
entropy of the topologically constrained polymers remains a
difficult challenge \cite{Nechaev}.  More recently, some insights
into the statistical properties of knots were developed based on
various scaling arguments
\cite{tube_inflation,Stasiak_book,3_5_uzla,Kardar1,Kardar2,Kardar5}.
As in other cases in soft condensed matter physics
\cite{deGennes_book,deGennes_Views}, scaling is the theory of
choice where more sophisticated approaches are too difficult.

The simplest question of polymer physics is that about the size
(say, root-mean-squared gyration radius) of a single macromolecule
in a dilute solution.  In general, this question is not understood
for the topologically constrained polymer ring.  As a first step
in this direction, it was recently shown \cite{3_5_uzla}
(confirming the earlier conjecture \cite{desCloizeaux}) that
topological constraints in the knot effectively lead to
self-avoidance on the large enough length scales.  As a result,
the size of a polymer ring with the topology of a trivial knot (an
unknot) scales as $R \sim \ell N^{\nu}$ even for the polymer with
negligible excluded volume.  Here, $N$ is the number of effective
segments, $\ell$ is the segment length, and $\nu \approx 0.588
\approx 3/5$ is the well known critical exponent of self-avoiding
walks \cite{deGennes_book}.  This result has been also extended
for the non-trivial knots using the aspect ratio of maximally
inflated tube, $p$, as a topological invariant
\cite{tube_inflation}.  Overall, the result of the work
\cite{3_5_uzla} reads
%
%\begin{equation} R \simeq \left\{ \begin{array}{lcr} \ell N^{1/2} p^{-1/6} & {\rm when} & N <
%p N_0   \\  \ell N^{3/5} p^{-4/15} N_0^{-1/10} & {\rm when} & N
%> p N_0
% \end{array} \right. \ . \end{equation}
%
%
\begin{equation} R \simeq \left\{ \begin{array}{lcr} \ell N^{1/2} p^{-1/6} & {\rm when} & N <
p N_{00}   \\  \ell N^{\nu} p^{ - \nu + 1/3}  N_{00}^{- \nu + 1/2}
& {\rm when} & N > p N_{00}
 \end{array} \right. \ . \label{eq:my_old_result}  \end{equation}
Here, $N_{00}$ is the characteristic length of knotting; it comes
from the observation \cite{Muthukumar,Deguchi} and mathematical
proof \cite{theorem} that the probability to have an unknot in a
phantom chain (which freely crosses itself) goes like
$e^{-N/N_{00}}$. Numerically, $N_{00} \approx 300$
\cite{Muthukumar,Deguchi}.  Thus, the theory predicts that $R \sim
N^{3/5}$, if $N$ is large enough, independently of the knot type.
This prediction was scrutinized using computer simulations
\cite{ShimamuraDeguchiDA,ShimamuraDeguchiNET,StasiakNeScience};
the most recent work \cite{StasiakNeScience} claims complete
agreement with the predicted scaling.

\begin{figure}
\centerline{\scalebox{0.35}{\includegraphics{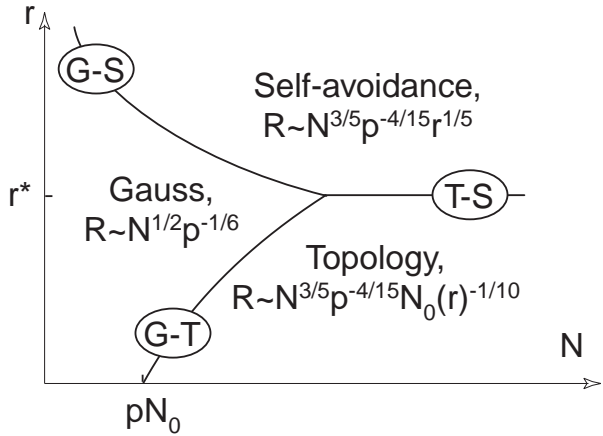}}}
\caption{Diagram of scaling regimes of a polymer ring.  $N$ is the
number of effective segments in the chain, $r$ is the
dimensionless thickness of the polymer (the ratio of the diameter
to the segment length), $p$ is the knot invariant - diameter to
the length ratio of maximally inflated tube
\protect\cite{tube_inflation}.  The scaling of polymer size for
every regime is indicated in the figure using the approximate
value of the critical exponent $\nu \approx 3/5$.  The cross-over
boundaries are as follows: the $G-S$ line between Gaussian and
self-avoidance regimes  $r^2 \simeq p/N$; the $G-T$ line between
Gaussian and topological regimes $N \simeq p N_0(r)$ (see Eq.
(\protect\ref{eq:N_nulevoe})); and the $T-S$ line between
topological and self-avoidance regimes $r = r^{\ast} \approx
0.03$. } \label{fig:diagram}
\end{figure}

In this paper, our goal is to study the interplay of topological
constraints and excluded volume interactions - two factors both of
which are inevitably present in any real polymer.  We imagine a
polymer ring of $N$ effective segments of the length $\ell$ and
diameter $d$ each.  We would like to emphasize that end closing
itself would not change the polymer size scaling, affecting only a
numerical prefactor.  This means, if we imagine a ring with
excluded volume, but phantom (for which all topological classes of
conformations are available, like in the presence of topological
enzyme \cite{Sikorav}), then the usual prediction
\cite{deGennes_book} holds:
\be R_g \sim \left\{ \begin{array}{lll} \ell N^{1/2} & {\rm when} & N < 1/r^2 \\
\ell N^{\nu} r^{2 \nu - 1} & {\rm when} & N > 1/r^2
\end{array} \right. \ . \label{eq:phantom} \ee
Thus, our goal if to generalize together the Eqs.
(\ref{eq:my_old_result}) and (\ref{eq:phantom}).   Our results are
summarized by the Fig. \ref{fig:diagram}, in which the diagram of
scaling regimes is presented in terms of variables $N$ and
$r=d/\ell$. The paper is organized around the discussion of this
diagram.

As an important input of our analysis, we shall rely on the
following insight gained by computer experiments
\cite{Klenin,Deguchi_excluded} regarding the probability of
knotting for the excluded volume polymer.  As one might have
guessed, excluded volume enhances the probability of a trivial
knot, somewhat suppressing all non-trivial knots.  What is much
more difficult to guess is that the probability of a trivial knot
remains exponential in $N$, with characteristic length very
sharply depending on the excluded volume parameter.  According to
\cite{Klenin,Deguchi_excluded},
\begin{eqnarray} p_{\rm unknot} (N,r) & \simeq & \exp \left[ -N /
N_0(r) \right] \ , \nonumber \\ N_0(r) & \simeq & N_{00} \exp
\left[ \beta r^{\upsilon} \right] \ , \nonumber \\ N_{00} &
\approx & 300 \ , \beta \approx 30 \ , \upsilon \approx 1 \ .
\label{eq:N_nulevoe}
\end{eqnarray}
In fact, the authors of the work \cite{Deguchi_excluded} found
that according to their numerical data, the value of $\upsilon$ is
slightly below unity, $\upsilon \approx 0.85$.  They also claimed
that the coefficient $A$ me depend on the model.  Neither of these
details is important for our purposes here, we will only use the
fact that the characteristic length of knotting, $N_0(r)$ is
sharply increasing with $r$.

We are now in a position to start discussing the scaling diagram,
Fig. \ref{fig:diagram}.  Let us begin with a trivial knot, or an
unknot, in which case $p$ (in the scaling sense) is of order unity
($p = 2 \pi$).

First of all, there is obviously a regime when neither excluded
volume nor topological constraints are important.  This range is
labeled as Gaussian in Fig. \ref{fig:diagram}, because chain size
in this regime scales as that of a Gaussian polymer, $R \sim
N^{1/2}$.  This happens if the conditions $N < 1/r^2$
(\ref{eq:phantom}) and $N< N_0(r)$ (\ref{eq:N_nulevoe}) are met
simultaneously.  In particular, the latter condition, according to
the formula (\ref{eq:N_nulevoe}) ensures that even for the phantom
chain the part of conformation space corresponding to all
non-trivial knots is very small; therefore, set of conformations
of phantom chain is practically the same as that of a trivial
knot.

When we cross over the $G-S$ boundary $N \simeq 1/r^2$, excluded
volume effect takes over, but topological constraints may still
remain unimportant if $N < N_0(r)$.  In this regime, labeled as
dominated by self-avoidance in Fig. \ref{fig:diagram}, chain size
is given by $R \simeq \ell N^{\nu} r^{2 \nu - 1}$.

Another regime is obtained from Gaussian if we cross over the
$N=N_0(r)$ line $G-T$, remaining at $N < 1/r^2$.  In this case, we
should imagine the chain as consisting of $N/N_0(r)$ blobs, of
$N_0(r)$ monomers each.  Inside each blob, as it length is smaller
than the characteristic knotting length, the topological
constraints are insignificant, as well as the excluded volume;
therefore, the blob size is about $R_{\rm blob} \simeq \ell
N_0(r)^{1/2}$.  On the other hand, the chain of blobs is
topologically constrained, and, therefore, according to
\cite{3_5_uzla}, we get for the chain size $R \simeq R_{\rm blob}
\left( N / N_0(r) \right)^{\nu} \simeq \ell N^{\nu} N_0(r)^{-\nu +
1/2} $.

At this stage, it is useful to check the self-consistency of our
result.  As we know \cite{tube_inflation,3_5_uzla}, the trivial
knot size results from the balancing of two entropic factors.  On
the one hand, chain has entropic elasticity $\sim R^2/N \ell^2$
which resists swelling.  On the other hand, topologically
constrained blobs resist interpenetration.  At the equilibrium
$R$, both entropic contributions are of the same order.  We must
now check that the second virial term corresponding to the
excluded volume interactions is smaller than either of the above
mentioned terms. Given that the second virial coefficient of
elongated (rod-like) segments is about $\ell^2 d$, the
self-consistency condition reads
\begin{equation} \ell^2 d \frac{N^2}{R^3} < \frac{R^2}{\ell^2 N} \
\label{eq:consist} . \end{equation}
Note, that this condition is equivalent to $R > R_{\rm Flory}$,
where $R_{\rm Flory} \simeq \ell N^{3/5} r^{1/5}$ is the Flory
estimate of polymer size without topological constraints
\cite{deGennes_book}.  Thus, this condition means simply that
excluded volume is unimportant if the chain swells more for
topological reasons than it would due to the excluded volume.
Simple algebra indicates that this condition is met if $N_0(r) <
1/r^2$, which is equivalent to $r < r^{\ast}$.  Numerically,
according to eq. (\ref{eq:N_nulevoe}), $r^{\ast} \approx 0.03$.

Finally, let us see what happens if $N_0(r) > 1/r^2$ and $N >
N_0(r)$.  Diagram Fig. \ref{fig:diagram} indicates that in this
case we cross over the $T-S$ boundary and enter again the same
self-avoidance regime which we discussed previously n terms of
crossing over the $G-S$ line.  Let us show why this is indeed the
same regime.  We should imgine once again that our chain consists
of $N/N_0(r)$ blobs, with $N_0(r)$ monomers in each blob.  In the
present case, since $N_0(r) > 1/r^2$, excluded volume is important
for each blob, so that the blob size is about $R_{\rm blob} \simeq
\ell N_0(r)^{\nu} r^{2 \nu - 1 }$.  Since the chain of blobs is
topologically constrained, overall chain size scales as $R \simeq
R_{\rm blob} \left( N / N_0(r) \right)^{\nu} \simeq \ell N^{\nu}
r^{2 \nu - 1 } $, which does not involve the topological quantity
of $N_0(r)$ and coincides with our result for the self-avoidance
dominated regime.

This completes the analysis for the case of an unknot (trivial
knot).

Let us now consider some non-trivial knot, characterized by the
length-to-diameter ratio of maximally inflated tube, also called
ideal knot representation.  The specific values of $p$ for many
knots were computed by Pieranskii et al \cite{Pieranskii}.
Following the ideology of \cite{tube_inflation,3_5_uzla}, we
imagine the polymer as confined in the maximally inflated tube
such that it is unknotted inside the tube.  This corresponds to
viewing the polymer as consisting of $p$ roughly spherical blobs,
with $N/p$ monomers in each blob.  Furthermore, since the tube is
maximally inflated, the blobs fill completely (in the scaling
sense) the entire volume of the coil, which means $R^3 \simeq p
R_{\rm blob}^3$, or
\begin{equation} R \simeq p^{1/3} R_{\rm blob} \ . \label{eq:dense_blobs}
\end{equation}
Therefore, all we have to do is to address all possible regimes of
the chain inside the blob.

The simplest case is when neither excluded volume nor topology are
of importance inside the blob.  This happens when both $N/p <
1/r^2$, which is below the $G-S$ line in Fig. \ref{fig:diagram},
and $N/p < N_0(r)$, which is above or to the left of the $G-T$
line.  In this regime every blob is Gaussian, yielding $R_{\rm
blob} \simeq \ell (N/p)^{1/2}$ and, according to formula
(\ref{eq:dense_blobs}),
\begin{eqnarray} R \simeq \ell N^{1/2} p^{-1/6} \ , {\rm when} \ N & < & p/r^2
\nonumber \\ {\rm and} \ N  & < & p N_0(r) \ . \end{eqnarray}
This is Gaussian regime of Fig. \ref{fig:diagram}.

When $N/p$ exceeds $1/r^2$, we cross over the $G-S$ line into the
regime dominated by self-avoidance.  In this case, the blob size
is about $R_{\rm blob} \simeq \ell (N / p)^{\nu} r^{2 \nu -1}$.
Therefore, formula (\ref{eq:dense_blobs}) yields
\begin{eqnarray} R \simeq \ell N^{\nu} p^{-\nu +1/3} r^{2 \nu -1} \ , {\rm when} \ N & > & p/r^2
\nonumber \\ {\rm and} \ r  & > & r^{\ast} \ .
\end{eqnarray}
Similar to the case of the trivial knot, this regime, which we
call self-avoidance dominated, is valid all the way from the $G-S$
boundary $N \simeq p/r^2$ to the $T-S$ boundary $r = r^{\ast}$.

Finally, if $r$ is small, but the blob length $N/p$ exceeds
$N_0(r)$, then blobs become topologically constrained.  This
means, it becomes important that the chain inside the tube is
unknotted, and so must be every blob.  In this regime, blob size
scales as $R_{\rm blob} \simeq \ell (N/p)^{\nu} N_0(r)^{-\nu +
1/2}$.  Plugging this into formula (\ref{eq:dense_blobs}) leads to
\begin{eqnarray} R \simeq \ell N^{\nu} p^{-\nu +1/3} N_0(r)^{-\nu +1/2} \ , {\rm when} \ N & > & p
N_0(r) \nonumber \\ {\rm and} \ r  & < & r^{\ast} \ .
\end{eqnarray}
This is topology dominated regime.

\begin{figure}
\centerline{\scalebox{0.35}{\includegraphics{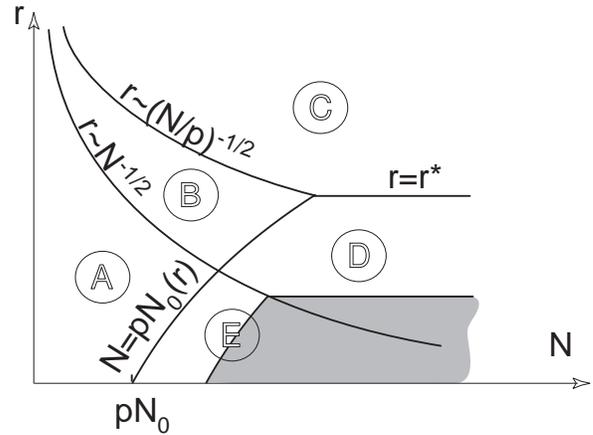}}}
\caption{Scaling regimes for the quantity $\xi$
(\protect\ref{eq:xi}). As in Fig. \protect\ref{fig:diagram}, the
results are shown using the approximate value of the critical
exponent $\nu \approx 3/5$.  Regime A: both topologically
constrained and phantom chains are Gaussian, $\xi \simeq p^{-1/6}
< 1 $.  Regime B: phantom polymer swells due to excluded volume,
while knot remains Gaussian: $\xi \simeq p^{-1/6} (N r^2)^{-1/10}
< 1 $.  Regime C: both phantom polymer and the knot swell due to
the excluded volume, $\xi \simeq p^{-4/15} < 1$.  Regime D: knot
swells because of topology, while phantom ring swells due to the
excluded volume, $\xi \simeq p^{-4/15} (N_0(r) r^2)^{-1/10}$.
Regime E: knot is swollen due to the topological constraint, while
phantom ring is Gaussian, $\xi \simeq p^{-4/15} \left( N / N_0(r)
\right)^{1/10}$. In the shaded area, which includes parts of the
regimes D and E, the topologically constrained polymer is more
expanded than the phantom (averaged of all knots). }
\label{fig:diagram2}
\end{figure}

It is instructive to summarize the results by looking at the ratio
of the given knot size $R$ to the size of a phantom ring polymer
(\ref{eq:phantom}), with the same length $N$, segment length
$\ell$, and excluded volume $r$:
\begin{equation} \xi = \frac{R_{\rm knot}(p)}{R_{\rm phantom}} \ . \label{eq:xi} \end{equation}
This quantity is useful, because $R_{\rm phantom}$ can be also
understood as the average size over all knot types, and such
quantity can be easily extracted from computer simulations data
\cite{ShimamuraDeguchiDA,ShimamuraDeguchiNET}.  Various regimes
for this quantity are outlined in the Fig. \ref{fig:diagram2},
where the cross-over lines of the scaling diagram Fig.
\ref{fig:diagram} are superimposed with the line $r \simeq 1 /
\sqrt{N}$ separating Gaussian and swollen regimes for the phantom
polymer.

Somewhat counterintuitive result is that there is a parameter
range (shaded in Fig. \ref{fig:diagram2}) in which topologically
constrained knot is more expanded than phantom average over all
knots.  This happens because for the phantom chain (with annealed
topological disorder) in this range more complex knots (with $p$
larger than the given value) are prevailing, they are more
compact, and they dominate the average $R_{\rm phantom}$.  This
result is at least qualitatively in agreement with the simulation
data \cite{ShimamuraDeguchiDA}.

Metzler et al \cite{Kardar1,Kardar2} recently reexamined the
possibility of "knot segregation," which was briefly discussed at
the end of \cite{tube_inflation} and later analyzed numerically
\cite{KatrichStasiak_segregation}.  The idea is that the entropy
of a knotted polymer may be maximaized by the conformations in
which knot is tightened within a short part of the chain, while
the rest of the polymer fluctuates as a free unknotted loop. While
the tendency to tighten the knot is quite strong for charged
polymers \cite{Kardar5}, for neutral system it is governed by
sub-linear in $N$ terms of free energy and, therefore, is very
sensitive to the details of structure and interactions.  In this
paper, we assumed that topological constraints are delocalized
along the polymer chain backbone.  In the context of our work,
this is justified for the following reasons.  First, this issue is
altogether irrelevant for the important case of the trivial knot,
which is the first case considered in this paper.  Second, for the
non-trivial knots, it is necessary to understand the delocalized
states in order to be able to push the question of knot
segregation beyond the slip-link model of \cite{Kardar2}.

It should also be noted that in this paper we always assumed knots
simple enough such that polymer chain is far from fully stretched
state in the tube.  The regime of such strong knotting requires
special analysis \cite{Rabin}.

To conclude, we presented the scaling theory determining the size
of a polymer ring with quenched topology of a certain knot.

\acknowledgments The author acknowledges useful discussion and
correspondence with M. Shimamura and T. Deguchi.

%\begin{references}


\begin{thebibliography}{}

\bibitem{Cozarelli} V. Rybenkov, N. Cozzarelli, A. Vologodskii
\emph{Proc. Nat. Ac. Sci.}, v. \textbf{90}, 5307, 1993.

\bibitem{Wang} S. Shaw, J. Wang \emph{Science}, v. \textbf{260},
533, 1993.

\bibitem{Rybenkov} V. Rybenkov, C. Ullsperger, A. Vologodskii, N. Cozzarelli
%Simplification of DNA topology below equilibrium values by type II topoisomerases
\emph{Science}, \textbf{277}, 690-693, 1997.

\bibitem{Kaufman} L. Kauffman \emph{knots and physics}, World
Scientific, Singapoor, 2001.

\bibitem{Maxim} A.V.Vologodskii, A.V.Lukashin, M.D.Frank-Kamenetskii
{\it Sov. Phys. JETP}, v. {\bf 67}, p. 1875, 1974;
A.V.Vologodskii, M.D.Frank-Kamenetskii  {\it Sov. Phys. Uspekhi},
v. {\bf 134}, p. 641, 1981;  A.V.Vologodskii {\it Topology and
Physics of Circular DNA}, CRC Press, Boca Raton, 1992.

\bibitem{Muthukumar} K.Koniaris, M.Muthukumar
% ``Knotedness in Ring Polymers,''
Phys. Rev. Lett., v. \textbf{66}, p. 2211-2214, 1991.
\bibitem{Deguchi} T.Deguchi, K.Tsurusaki
% ``Universality of Random Knotting,''
Phys. Rev. E, v. \textbf{55}, n. 5, p. 6245-6248, 1997.

\bibitem{Deutch} J.M.Deutsch
%``Equilibrium size of large ring molecules''
Phys. Rev. E, v. \textbf{59}, n. 3, p. R2539, 1999.

\bibitem{Nechaev} S.Nechaev \emph{Statistics of knots and entangled
random walks}, (World Scientific: Singapore, 1996)

\bibitem{tube_inflation} A.Yu.Grosberg, A.Feigel, Y.Rabin
% ``\emph{ Flory-Type Theory of a Knotted Ring Polymer\/}''
Physical Review E, v. \textbf{54}, n. 6, p. 6618-6622, 1996.

\bibitem{Stasiak_book} V.Katrich, A.Stasiak, L.Kauffman, Editors Ideal Knots,
World Scientific, 1999.

\bibitem{3_5_uzla} A. Grosberg %``{\it Critical Exponents for Random Knots},''
\emph{Phys. Rev. Let.},  \textbf{85}, 3858-3861, 2000.

\bibitem{Kardar1} R. Metzler, A. Hanke, P. Dommersnes, Y. Kantor, M. Kardar % Equilibrium shapes of flat knots
\emph{Phys. Rev. Lett.}, \textbf{88}, 188101, 2002.
\bibitem{Kardar2} R. Metzler, A. Hanke, P. Dommersnes, Y. Kantor, M. Kardar
\emph{Phys. Rev. E}, \textbf{65}, 061103, 2002. %Tightness of slip-linked polymer chains
%\bibitem{Kardar3} O. Farago, Y. Kantor, M. Kardar  cond-mat/0205111 %Pulling Knotted Polymers
%\bibitem{Kardar4} R. Metzler, Y. Kantor, M. Kardar  cond-mat/0206057 %Force-Extension Relations for Polymers with Sliding Links
\bibitem{Kardar5} P. Dommersnes, Y. Kantor, M. Kardar  cond-mat/0207276 %Knots in Charged Polymers

\bibitem{deGennes_book} P.G. de Gennes \emph{Scaling Concepts in Polymer
Physics}, Cornell Univ.Press, Ithaca, NY, 1979.
\bibitem{deGennes_Views} P.G. de Gennes \emph{Simple Views on Condensed Matter}, World Scientific, Singapoor, 1992.

\bibitem{desCloizeaux} J. des Cloizeaux
% ``Ring polymers in solution: topological effects,''
\emph{J.Physique Lettres}, v. \textbf{42}, p. L433, 1981.

\bibitem{ShimamuraDeguchiDA} M. Shimamura, T. Deguchi
% Gyration radius of a circular polymer under a topological constraint with exluded volume
\emph{Phys. Rev. E}, v. \textbf{64}, 020801, 2001.

\bibitem{ShimamuraDeguchiNET} M. Shimamura, T. Deguchi
% Finite-size and asymptotic behaviors of the gyration radius of knotted cylindrical self-avoiding polygon
\emph{Phys. Rev. E}, v. \textbf{65}, 051802, 2002.

\bibitem{StasiakNeScience} A. Dobay et. al., private communication.

\bibitem{theorem} D.W.Sumners, S.G.Whittington
% ``Knots in Self-Avoiding Walks''
\emph{ Journal of Physics A: Math.\& Gen.}, v. \textbf{21}, p.
1689, 1988; N.Pippenger
%``Knots in Random Walks''
\emph{ Disc. Appl. Math.}, v. \textbf{25}, 273, 1989.

\bibitem{Sikorav} Sikorav J.-L., B. Duplantier, G. Jannink \& Y. Timsit.
% DNA crossovers and type II DNA topoisomerases: A thermodynamical study.
\emph{J. Mol. Biol.}, v. \textbf{284}, p. 1279-1287, 1998

\bibitem{Klenin} K. Klenin, A. Vologodskii, V. Anshelevich, A. Dykhne, M. Frank-Kamenetskii
\emph{J. Biomol. Struct \& Dyn.}, v. \textbf{5}, 1173, 1988.

\bibitem{Deguchi_excluded} M. Shimamura, T. Deguchi
%Characteristic length of random knotting for cylindrical self-avoiding polygons
\emph{Phys. Lett. A}, v. \textbf{274}, 184, 2000.

\bibitem{Pieranskii} P.Pieranskii et al, contribution in the book
\protect\cite{Stasiak_book}.

\bibitem{KatrichStasiak_segregation} V. Katritch, W. Olson, A. Vologodskii, J. Dubochet, A. Stasiak
\emph{Phys. Rev. E}, \textbf{61}, 5545-5549, 2000.  % Tightness of random knotting

\bibitem{Rabin} A. Grosberg, Y. Rabin, unpublished result.

\end{thebibliography}
\end{document}